# Synthesis and Characterization of the Ternary Nitride Semiconductor Zn$_2$VN$_3$: Theoretical Prediction, Combinatorial Screening and Epitaxial Stabilization

Siarhei Zhuk[a], Andrey A. Kistanov[b], Simon C. Boehme[a,c], Noémie Ott[a], Fabio La Mattina[a], Michael Stiefel[a], Maksym V. Kovalenko[a,c], Sebastian Siol[a,*]

a) Empa – Swiss Federal Laboratories for Materials Science and Technology, 8600 Dübendorf, Switzerland

b) Nano and Molecular Systems Research Unit, University of Oulu, 90014 Oulu, Finland

c) Laboratory of Inorganic Chemistry, Department of Chemistry and Applied Bioscience, ETH Zürich, 8093 Zürich, Switzerland

*E-mail: sebastian.siol@empa.ch

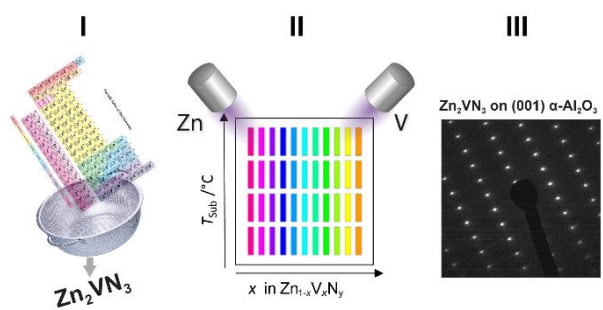

**Graphical abstract:** Prediction and synthesis of the novel ternary nitride Zn$_2$VN$_3$

**Abstract**

Computationally guided high-throughput synthesis is used to explore the Zn-V-N phase space, resulting in the synthesis of a novel ternary nitride $Zn_2VN_3$. Following a combinatorial PVD screening, we isolate the phase and synthesize polycrystalline $Zn_2VN_3$ thin films with wurtzite structure on conventional borosilicate glass substrates. In addition, we demonstrate that cation-disordered, but phase-pure (002)-textured, $Zn_2VN_3$ thin films can be grown using epitaxial stabilization on α-$Al_2O_3$ (0001) substrates at remarkably low growth temperatures well below 200 °C. The structural properties and phase composition of the $Zn_2VN_3$ films are studied in detail using XRD and (S)TEM techniques. The composition as well as chemical state of the constituent elements are studied using RBS/ERDA as well as XPS/HAXPES methods. These analyses reveal a stoichiometric material with no oxygen contamination, besides a thin surface oxide. We find that $Zn_2VN_3$ is a weakly-doped p-type semiconductor demonstrating broadband room-temperature photoluminescence spanning the range between 2 eV and 3 eV. In addition, the electronic properties can be tuned over a wide range via isostructural alloying on the cation site, making this a promising material for optoelectronic applications.

# 1. Introduction

The discovery of new functional materials remains paramount for the continuous technological advancements. [1] For instance, nitrides are essential for a number of electronic applications, ranging from transistors and solid-state lighting to diffusion barrier layers, owing to the unique combination of electronic, optical and structural properties.[2–5] For optoelectronic applications, ternary transition- and post-transition metal nitrides are particularly promising, due to their adjustable band gaps and favorable transport properties. [6,7]

Strikingly, nitrides remain a rather underexplored class of materials [8–11] in part, due to their challenging synthesis. For example, as the formation of the corresponding oxide phases is energetically often more favorable the synthesis of nitrides often requires particularly clean or highly energetic synthesis environments.[12,13] The enormous potential of this phase space was recently demonstrated in a comprehensive high-throughput computational study by Sun *et al.*, through the prediction of hundreds of new stable and metastable ternary metal nitrides.[8] Such first-principles density functional theory (DFT) calculations of material' properties, such as their electronic structure and thermodynamic stability, can significantly facilitate the experimental synthesis and discovery of materials with desired properties for specific applications.[14–18]

Ternary zinc nitrides (Zn-Me-N), in particular Zn-IV-$N_2$ materials, have attracted much attention in recent years.[19,20] For instance, $ZnGeN_2$, which exhibits similar optical and structural properties compared to GaN, has been discussed as an attractive candidate for the fabrication of light-emitting diodes.[21,22] Another prominent example is $ZnSnN_2$. After its theoretical prediction in 2008, this material has been studied extensively for solar energy conversion owing to its suitable optical and electronic properties.[23–26] The number of reported new functional nitrides has been steadily increasing. Arca *et al.* demonstrated the novel metastable $Zn_2SbN_3$ with a direct band gap of 1.55 eV and wurtzite crystal structure.[9] Other novel wurtzite-derived ternary Zn nitrides include $Zn_2NbN_3$ [27] and $Zn_3MoN_4$.[28] In addition, Woods-Robinson *et al.* reported metastable $Zn_xZr_{1-x}N_2$ alloyed thin films of rocksalt and hexagonal boron-nitride crystal structure.[29] In many of these works combinatorial physical vapor deposition (PVD) was successfully employed, which showcases the potential of high-throughput combinatorial techniques for screening of the complex synthesis phase spaces associated with multinary nitride material systems.[30–33]

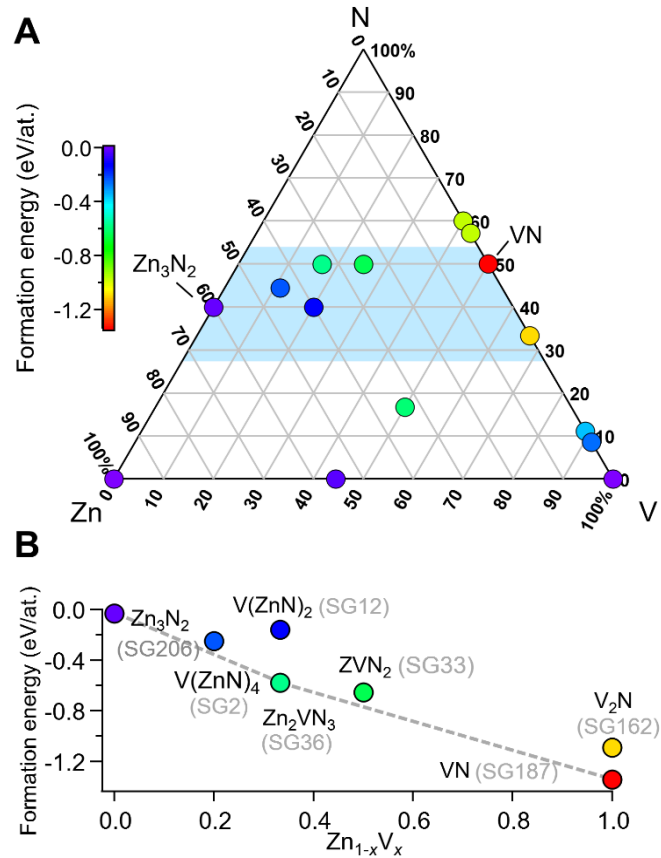

**Figure 1.** Formation energy of stable and metastable, binary, and ternary compounds for Zn-V-N phase space. (a) Ternary phase diagram of Zn-V-N. Phases with an anion-to-cation ratios close to 1 are highlighted in blue. (b) Convex hull construction of Z-V-N compounds with anion/cation ratio close to unity, including the potentially stable $Zn_2VN_3$ as well as selected metastable compounds.[34]

In this work, we use a combination of computational prediction and PVD synthesis to take a closer look at the Zn-V-N phase space. In this phase space a number of interesting computationally predicted compounds remain to be demonstrated experimentally. **Figure 1a** shows a ternary phase diagram of the Zn-V-N system with calculated formation energies for various binary and ternary compounds.[34] $Zn_2VN_3$ compound with a band gap of ~1.3 eV is predicted in this phase space as well as metastable $ZnVN_2$, $Zn_2VN_2$ and $Zn_4VN_4$.[34] **Figure 1b** shows a convex hull construction for different zinc vanadium nitrides, with an anion-to-cation ratio close to unity. It is noteworthy that $Zn_2VN_3$ and $ZnVN_2$ are predicted to be orthorhombic, while $Zn_2VN_2$ and $Zn_4VN_4$ are expected to be monoclinic and triclinic, respectively.[34] On the other hand, in the computational screening of Hinuma *et al.* $Zn_2VN_3$ is predicted to exhibit wurtzite crystal structure.[35] To our knowledge, the synthesis of these compounds has not yet been reported. In this work, this region of the Zn-V-N materials system is covered using a combinatorial reactive PVD screening approach. The study results in the discovery and synthesis of a novel wurtzite $Zn_2VN_3$ phase. We demonstrate how phase-pure $Zn_2VN_3$ can be synthesized on large areas using reactive co-sputtering on diverse substrates. The structure, composition, electronic and optical properties of $Zn_2VN_3$ thin films are discussed in detail.

## 2. Methods

### 2.1 Synthesis of combinatorial sample libraries

1.1 mm thick borosilicate glass substrates of 50.8 mm x 50.8 mm size were ultrasonically cleaned in acetone and ethanol for 5 min each before loading into the sputter chamber (AJA 1500F). The thin films were synthesized via reactive radio-frequency (RF) co-sputtering from Zn (99.995% purity, 2" diameter) and V (99.9% purity, 2" diameter) targets. The deposition was carried out from unbalanced magnetrons in a closed-field magnetic configuration in mixed Ar and $N_2$ atmosphere. Combinatorial sample libraries of Zn-V-N thin films were grown at a process pressure of 1 Pa with Ar and $N_2$ flows of 12 sccm and 18 sccm, respectively. $N_2$ gas was introduced directly into the chimneys of the sputter guns, facilitating dissociation and ionization of the reactive gas and thus increasing its chemical potential. The substrate was static during the deposition to obtain a composition gradient across the combinatorial sample libraries containing 45 samples located in 5 rows and 9 columns. A custom-designed sample holder was used to achieve a temperature gradient on the substrate. The deposition temperature was maintained at 114 °C, 157 °C and 220 °C for a bottom, middle and top rows of the library, respectively. More details are available in the Supplementary Information.

## 2.2 Characterization of combinatorial sample libraries

Automated mapping analysis of structural properties was carried out using θ-2θ mapping in a Bruker D8 X-ray diffraction (XRD) system with Bragg Brentano geometry using Cu Kα radiation and a Ni filter. For selected samples, grazing-incidence XRD (GI-XRD) at an incident angle of ~1.5° was carried out. An XRD refinement was performed using MAUD software after obtaining the instrument parameters using a corundum reference sample. [36] For orthorhombic $Zn_2VN_3$ the conventional unit cell from the DFT calculations was used as a starting structure. For wurtzite $Zn_2VN_3$, a ZnN wurtzite structure (mp-971911) was modified using a cation-site occupancy of 0.6667 Zn and 0.3333 V. Mapping of the elemental ratio of Zn and V in the bulk of the films was performed using a Bruker M4 Tornado X-ray fluorescence (XRF) system with a Rh X-ray source. For selected samples, the bulk composition of the film was determined by Rutherford backscattering spectrometry (RBS) and elastic recoil detection analysis (ERDA) at the 1.7 MV Tandetron accelerator facility of the Laboratory of Ion Beam Physics at ETH Zurich. The RBS measurements were performed using a 2 MeV 4He beam and the collected RBS data were simulated using the RUMP software for the composition of Zn and V. For the ERDA analysis a 13 MeV 127I beam was used under incidence angle of 18° and the scattered recoils were identified by the combination of a time-of-flight spectrometer with a gas ionization chamber. Atomic fractions of the light elements C, N and O have been calculated from ERDA mass spectra approximately 50 nm below the film surface (to exclude the visible surface oxidation). X-ray photoelectron spectroscopy (XPS) mapping analysis was performed in a PHI Quantera system using monochromatic Al Kα radiation. The analysis was conducted at a pressure ~1-2·$10^{-6}$ Pa, whereas the electron beam was generated at a power of 50 W and a voltage of 15 kV. Hard X-ray photoelectron spectroscopy (HAXPES) analysis was carried out on selected samples in a PHI Quantes using a Cr Kα X-ray radiation source. The linearity of the energy scale and work function of the analyzer were calibrated using Au as well as Cu reference samples. Charge neutralization was performed using a low-energy electron source. The binding energy scale was referenced to the main component of adventitious carbon at 284.8 eV resulting in a typical inaccuracy of +/- 0.2 eV, which does not affect the values for the modified Auger parameter. Peak fitting was performed after Shirley background subtraction using Voigt profile with GL ratios of ~30 for the XPS as well as ~70 for the HAXPES measurements. Transmittance and reflectance spectra were measured with a UV-Vis-NIR spectrophotometer (Shimadzu UV-3600) equipped with an integrating sphere. A home-built four-point probe system was employed to perform mapping of the sheet resistance. In addition, a surface profilometer (Bruker Dektak XT) was used to measure the film thickness. The obtained experimental data sets were analyzed using custom written routines in IGOR Pro based on the framework of COMBIgor.[37] Photoluminescence (PL) spectra and time-resolved PL kinetics were acquired at room temperature in ambient conditions, using a fluorescence spectrometer (Picoquant FluoTime 300) equipped with a 355 nm pulsed laser and a time-correlated single photon counting unit. The employed laser repetition rate was 80 MHz and all PL spectra and kinetics have been corrected for dark counts and detector sensitivity. Microstructural analysis was performed using scanning transmission electron microscopy (S/TEM) combined with energy-dispersive X-ray spectroscopy (EDS). Imaging was performed using a JEOL 2200FS instrument, operating at 200 kV, fitted with an on-axis bright-field (BF) detector and a high-angle annular dark-field (HAADF) detector. Electronic transport measurements were performed on a single phase $Zn_2VN_3$ sample deposited on sapphire. Physical properties measurement system (PPMS, Quantum Design) was used for the temperature (50 to 300K) and magnetic field control (1, 2, and 3 T). Conductivity and Hall coefficient were acquired in typical Hall-bar geometry with four-probe schema. More details are provided in the supporting information.

## 2.3 Density functional theory-based calculations

The initial structures of $Zn_2VN_3$ and $ZnVN_2$ were created based on the primitive unit cell structures available in the Materials Project database (IDs mp-1029262 and mp-1246720, respectively).[34] The calculations were performed using the plane-wave method as implemented in the Vienna *ab initio* simulation package (VASP).[38] The Perdew–Burke–Ernzerhof (PBE) exchange–correlation functional under the generalized gradient approximation (GGA) and the Heyd–Scuseria–Ernzerhof (HSE) hybrid functionals were used for geometry optimization and electronic structure calculations, respectively.[39,40] Geometry optimization was stopped when the atomic forces and total energy values were smaller than 0.01 eV Å$^{-1}$ and 10$^{-6}$ eV, respectively. The first Brillouin zone was sampled with a 9×9×9 *k*-mesh grid. The kinetic energy cut-off was set to 450 eV. Periodic boundary conditions were applied in all three spatial directions.

## 3. Results and discussion

To support the experimental synthesis efforts, DFT calculations on selected phases in the Zn-V-N phase space were carried out using HSE hybrid functionals. The primitive unit cell structure of $Zn_2VN_3$ used for the calculations is presented in left panel of **Figure 2a**. In addition, a conventional cell structure of $Zn_2VN_3$ converted from the optimized primitive cell is presented in right panel of **Figure 2a**. The structure of $Zn_2VN_3$ was found to be orthorhombic with the space group 36 $Cmc2_1$ and the lattice constants a = 5.6 Å, b = 9.57 Å and c = 5.28 Å. The calculated energy band diagram is shown in **Figure 2c,** with an indirect bandgap of 2.23 eV and direct bandgap of 2.35 eV. In addition, it has been found that $Zn_2VN_3$ is a p-type semiconductor with the Fermi level located slightly above the valence band edge (see **Figure 2c**).

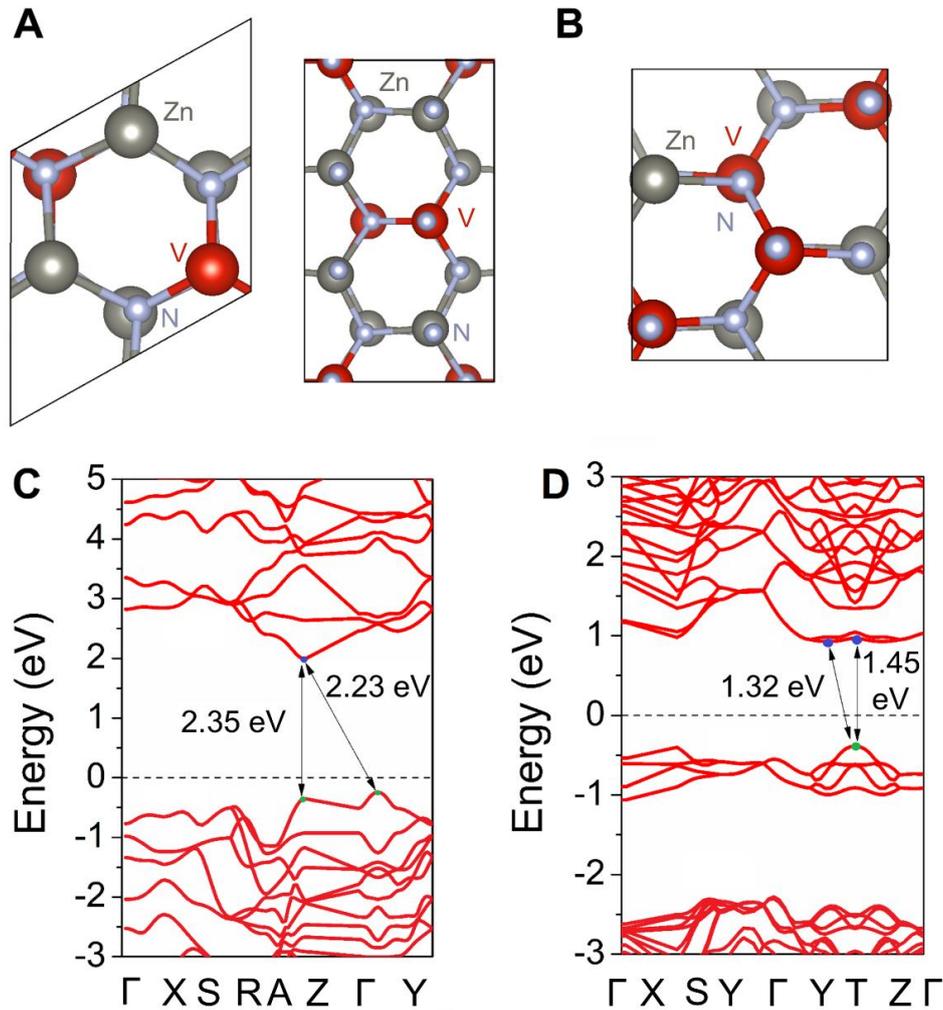

**Figure 2**. (a) Primitive (left) and conventional (right) unit cells of $Zn_2VN_3$. (b) primitive unit cell of $ZnVN_2$. Both materials are predicted to crystallize in an ordered orthorhombic structure. Energy band diagram obtained using HSE functionals for (c) $Zn_2VN_3$ and (d) $ZnVN_2$.

**Figure 2b** shows the crystal structure of $ZnVN_2$. This phase is stabilized as an orthorhombic crystal with the space group 33 $Pna2_1$ and the lattice constants a = 5.25 Å, b = 5.57 Å, and c = 6.42 Å. The calculated energy band diagram is shown in **Figure 2d**. The simulation predicted that $ZnVN_2$ is a p-type semiconductor with an indirect and direct bandgap of 1.32 eV and 1.45 eV, respectively. The electronic band structure of $ZnVN_2$ shows mid-gap states with small dispersion that are attributed to the V-d states caused by tetravalent V in the compound (see **Figure 2d**).

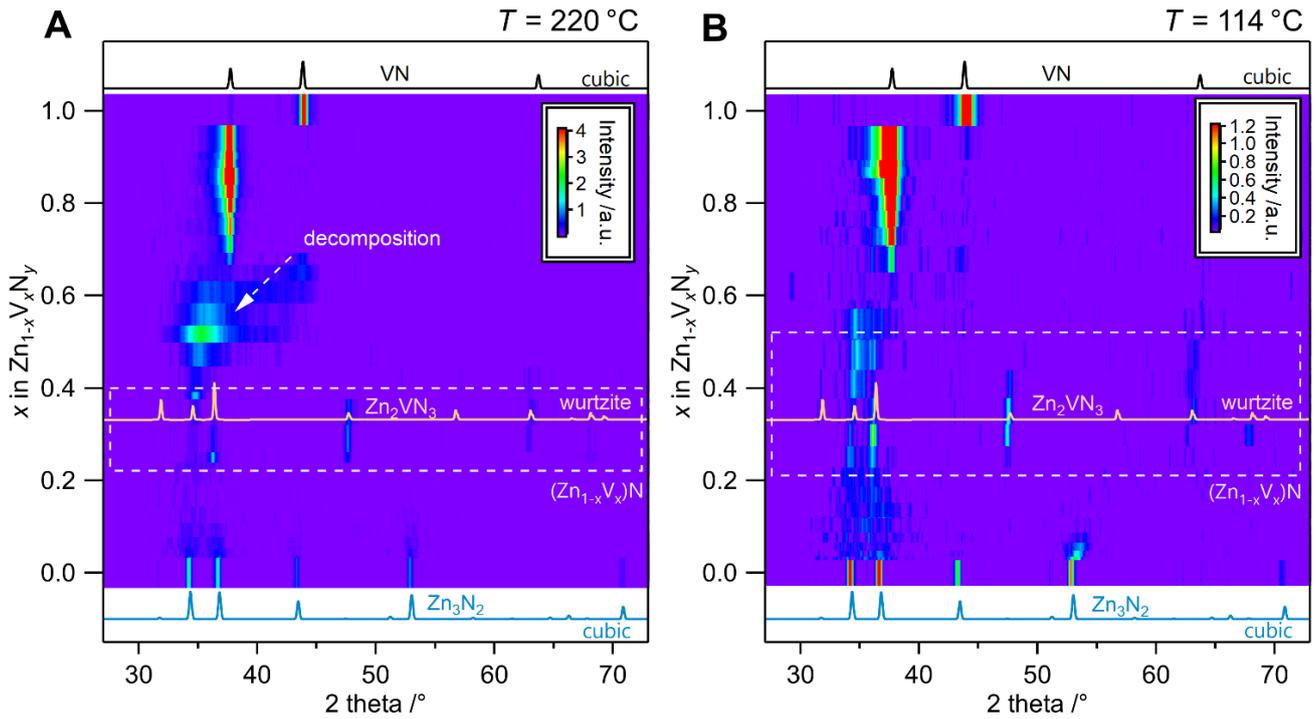

**Figure 3.** Combined XRD and XRF color map of Zn-V-N thin films synthesized at substrate temperatures of (a) 220 °C and (b) 114 °C. XRD patterns for $Zn_3N_2$ and VN [34] are given for reference as well as a simulated XRD pattern of $Zn_2VN_3$ in wurtzite structure.

Based on the predictions from DFT calculations, a phase-screening of the Zn-V-N phase space was carried out over the entire compositional range from $0 \leq x \leq 1$ in $Zn_{1-x}V_xN_y$ using XRD mapping on combinatorial sample libraries prepared by reactive co-sputtering. **Figure 3** shows a false-color plot of the XRD patterns for samples deposited at substrate temperatures of 220 °C as well as 114 °C as a function of composition and diffraction angle. XRD patterns for $Zn_3N_2$ and VN [34] are given for reference. Besides the expected stable compounds $Zn_3N_2$ and VN at the end of composition range, i.e. for $x = 0$ and $x = 1$, respectively, an additional phase can be observed at intermediate alloying concentrations. A more detailed XRD analysis reveals that in this compositional range $Zn_{1-x}V_xN_y$ crystallizes in wurtzite structure, rather than of the computationally predicted orthorhombic structure (see **Figure 4**). The major peak at 34.5° gradually shifts with increasing V content, which can be explained by a change in lattice parameter due to isostructural alloying on the cation site (see **Figure 3b**).

It is noteworthy that the region of stability for this phase is larger for lower deposition temperatures. For 114 °C it ranges approximately from $x$ = 0.2 to $x$ = 0.5, whereas for the deposition at 220 °C the formation of precipitates can be observed at V concentration as low as $x$ = 0.4 (see **Figure 3a**). According to DFT predictions, $ZnVN_2$ has a formation energy of -0.653 eV/atom while its energy above hull is 0.117 eV/atom [34] (see **Figure 1b**). This metastable nature of $ZnVN_2$ may be a possible reason for the observed decomposition at higher temperatures.

Following the combinatorial phase screening, the novel $Zn_2VN_3$ phase was successfully isolated on a glass substrate of 50.8 mm x 50.8 mm. Note that substrate rotation was used during this deposition process. **Figure S1** shows XRD patterns of all 45 samples for this sample library, demonstrating a good uniformity of the deposited thin film across a large area substrate. The predicted orthorhombic structure and the observed wurtzite phase are structurally closely related and exhibit similar XRD patterns. In order to unambiguously identify the structure of the synthesized phase, individual XRD analysis was carried out on selected points of this library to study the structural properties of $Zn_2VN_3$ in detail. A structural refinement based on GI-XRD analyses was performed using the MAUD software package. Using the calculated structure from DFT calculation the lattice parameters were refined to a = 5.623 Å, b= 9.719 Å and c = 5.191 Å. For the wurtzite phase a wurtzite ZnN structure was used as a starting point and altered to a random cation-site occupancy consisting of 66.67% Zn as well as 33.33% V. The structure was refined in terms of lattice parameters as well as atomic positions, resulting in a wurtzite lattice with the dimensions a = b = 3.243 Å and c = 5.191 Å, corresponding to a c/a ratio of 1.60. In addition, the refinement revealed average grain sizes on the order of 20 – 30 nm. Using these parameters as well as instrument broadening XRD patterns for the two structures were simulated.

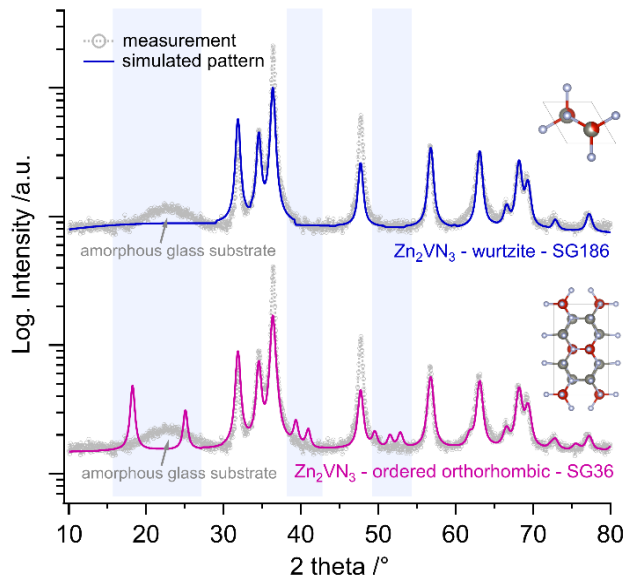

**Figure 4.** GI-XRD measurements of $Zn_2VN_3$ compared to simulated XRD patterns of orthorhombic $Zn_2VN_3$ (SG36) as well as wurtzite $Zn_2VN_3$ (SG186). The shaded regions highlight characteristic reflections of the ordered orthorhombic structure, which are not observed in the measurement.

**Figure 4** shows a comparison of the measured pattern of $Zn_2VN_3$ to the simulated pattern for each structure. A difference in peak height of some reflections is observed for both structures, which can be explained by a preferential out-of-plane orientation of the films (see **Figure 5C**) as well as the continuous change of the diffraction plane during the GI-XRD analysis. However, the pattern and peak positions are nicely matched by the simulated wurtzite pattern. The ordered orthorhombic structure on the other hand exhibits additional peaks at lower 2 theta values, which are not present in either the measured pattern further confirming the formation of a disordered wurtzite $Zn_2VN_3$ phase. To fully account for the effect of texture on the absence of these reflections the results were confirmed in additional XRD measurements at different tilts of the substrate (see **Figure S2**).

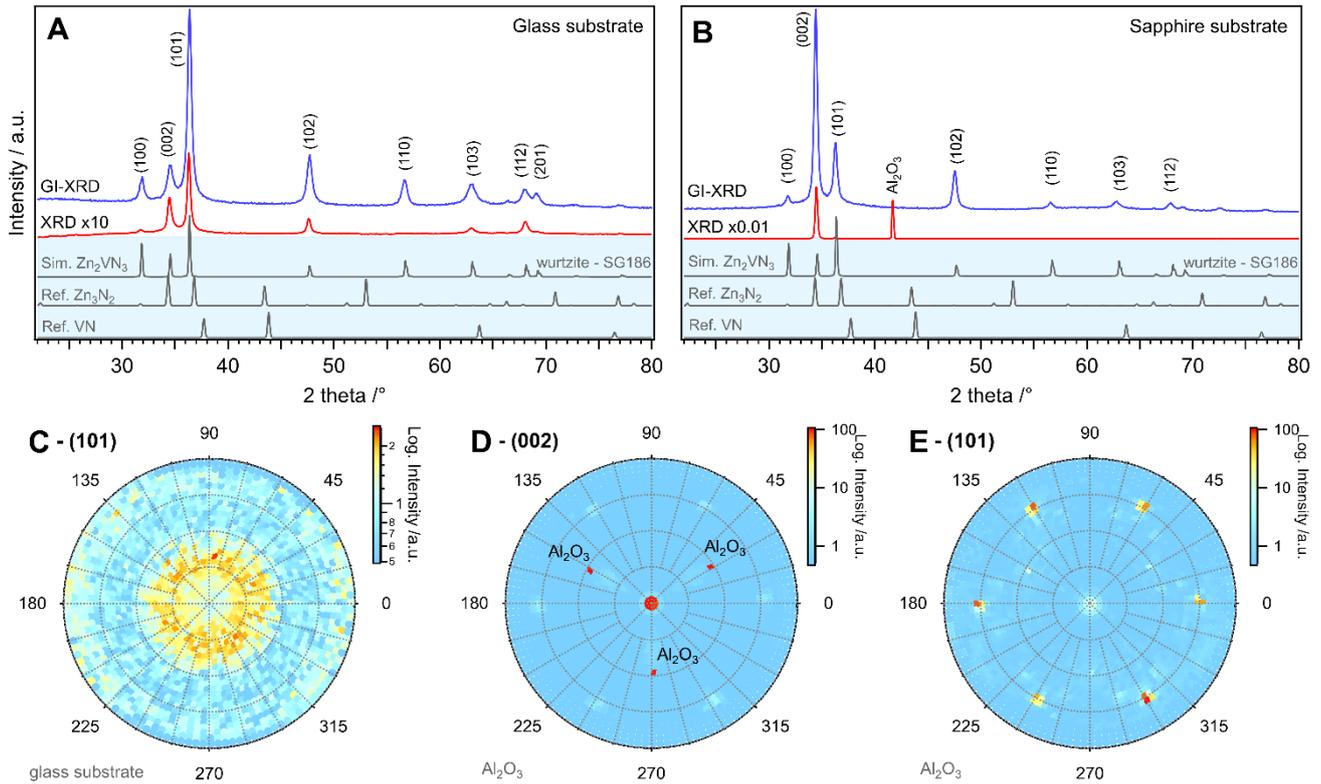

**Figure 5.** XRD and GI-XRD characterization of $Zn_2VN_3$ thin films deposited on (a) glass and (b) sapphire substrates. XRD pole figures of $Zn_2VN_3$ thin films: (c) (101) pole figure for a $Zn_2VN_3$ thin films on glass, (d), (e) (002) and (101) pole figures of a $Zn_2VN_3$ thin film grown on a sapphire substrate.

To increase the crystallinity of the material, a suitable substrate for epitaxial stabilization was selected. $Al_2O_3$ in (0001) orientation was chosen for its favorable minimal coincident interface area of 161.5 Å$^2$.[34,41] **Figure 5** shows a comparison of the structure and texture of $Zn_2VN_3$ films grown on glass as well as $Al_2O_3$ (0001) substrates. Shown are GI-XRD measurements as well as measurements in standard Bragg-Brentano geometry to highlight the differences in preferential orientation. In addition, pole figure analyses was performed to investigate the texture in more details. **Figure 5a** as well as **Figure 5c** show the XRD analysis of the thin films deposited on glass. The films exhibit a preferential (101) out-of-plane orientation as well as a rather low crystallinity. **Figure 5b** shows diffraction patterns of a $Zn_2VN_3$ thin film on a $Al_2O_3$ (0001) substrate. Along with an increase in crystallinity, the film exhibits a pronounced texture. Pole figure measurements of the (002) and (101) reflections indicate a preferential out-of-plane and in-plane orientation, as demonstrated in **Figures 5d** and **5e**, respectively. As shown in **Figure 5e**, six symmetrical peaks are visible with a weak central peak showing a 6-fold rotational symmetry – in good agreement with the formation of a disordered wurtzite structure. The difference in experimentally observed and theoretically predicted crystal structure is a direct result of pronounced cation-disorder in the material. Recently, Melamed *et al.* observed a similar effect in $ZnGeN_2$ thin films grown on $Al_2O_3$ substrate. Here the authors attributed crystallization of $ZnGeN_2$ in a wurtzite-type structure, instead of the theoretically predicted orthorhombic one, to a high degree of cation disorder.[42] In summary it can be concluded that both $Zn_2VN_3$ deposited on glass as well as sapphire substrates appear to be phase-pure with no visible impurities while the epitaxial stabilization on $Al_2O_3$ (0001) increases the crystallinity and texture of the material significantly.

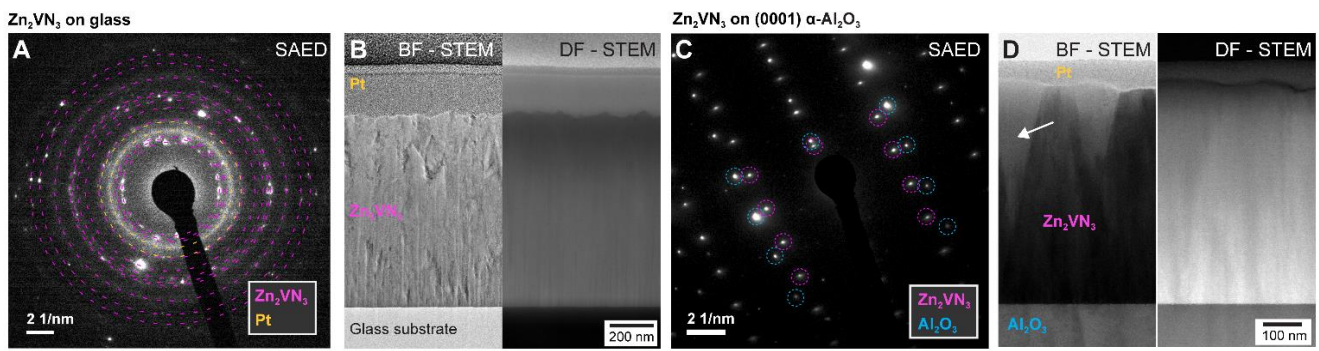

**Figure 6**. Microstructural analysis of $Zn_2VN_3$ on glass and $Al_2O_3$: Shown are selected area electron diffraction (SAED) (a,c), as well as bright field (BF) and dark (DF) field STEM micrographs (b,d). The electron diffraction pattern for films on glass (a) shows a polycrystalline structure and is well matched with the calculated reference (wurtzite $Zn_2VN_3$, SG186). Electron diffraction of $Zn_2VN_3$ on Al2O3 in the interface region demonstrates the epitaxial stabilization (c). STEM measurements on these films reveal grains of a different orientation closer to the surface of the film (d).

To perform a microstructural analysis of $Zn_2VN_3$, confirm the results of the XRD screening and reveal possible inhomogeneities, cross-sectional scanning transmission electron microscopy (STEM) analysis was conducted. **Figures 6a-d** show selected area electron diffraction (SAED) as well as bright field (BF) and dark (DF) field STEM micrographs for $Zn_2VN_3$ thin films grown on glass and $Al_2O_3$, respectively. The samples were covered with Pt prior to the TEM lamella preparation. The SAED pattern of $Zn_2VN_3$ on glass shown in **Figure 6a** matches the calculated diffraction pattern well and confirms the results from XRD analysis. The cross-sectional STEM BF image reveals a compact polycrystalline film. The vertical streaks are likely products of ion beam damage during the preparation of the TEM lamella. Most importantly, the DF image shows no significant lateral changes in intensity, which would be indicative of the formation of precipitates or secondary phases. A darkening is observed at the film's surface, which points to the formation of a surface oxide. In addition, the EDS map shows a very homogenous distribution of the film-forming elements (see **Figure S3b**). Measurements on $Zn_2VN_3$ grown on sapphire confirm the epitaxial growth. SAED measurements in the interface region reveal the closely matched structure of $Al_2O_3$ and $Zn_2VN_3$ (see **Figure 6c**). The orientation of the pattern is maintained throughout the entire specimen. The only exception are regions closer to the films surface (see **Figure S3d**) beyond a critical film thickness of ~ 200 nm. Here, STEM micrographs also show a change in contrast indicating the presence of differently oriented grains. These findings are in line with the pole figure analysis, which also revealed some (101) out-of-plane oriented grains (see **Figure 5e**). We can therefore conclude that $Zn_2VN_3$ is epitaxially stabilized on $Al_2O_3$ (0001). Moreover, the $Zn_2VN_3$ thin films investigated in this work exhibit a disordered wurtzite-like structure leading to the aforementioned 6-fold in-plane rotational symmetry.

The phase purity of $Zn_2VN_3$ samples was further assessed using Rutherford backscattering (RBS) analysis as well as elastic recoil detection analysis (ERDA) (see **Figure S4**). RBS measurements show that the films are almost perfectly stoichiometric with a Zn:V:N ratio of 2 : 1.02 : 3. ERDA analysis was further performed at a depth of approximately 50 nm, confirming a very low degree of oxygen contamination of well below 1 at.%. This very low concentration of oxygen was measured after storage in air for several days, indicating that the material is stable at ambient conditions.

The findings from RBS are in line with the results from photoelectron spectroscopy (see **Figure 7**). Analyzing nitrides using surface-sensitive techniques is often challenging due to the presence of surface oxides as well as adsorbates after exposure of the samples to air. To probe the chemical state and electronic structure below this surface layer, we used a combination of soft X-ray photoelectron spectroscopy (XPS) and hard X-ray photoelectron spectroscopy (HAXPES) using Al kα and Cr kα radiation, respectively. Depending on the kinetic energy of the excited electrons, the HAXPES measurement in $Zn_2VN_3$ can reach probing depths of up to 24 nm.[43] The calculated probing depth for each element are given in the Supporting Information in **Table S1**.

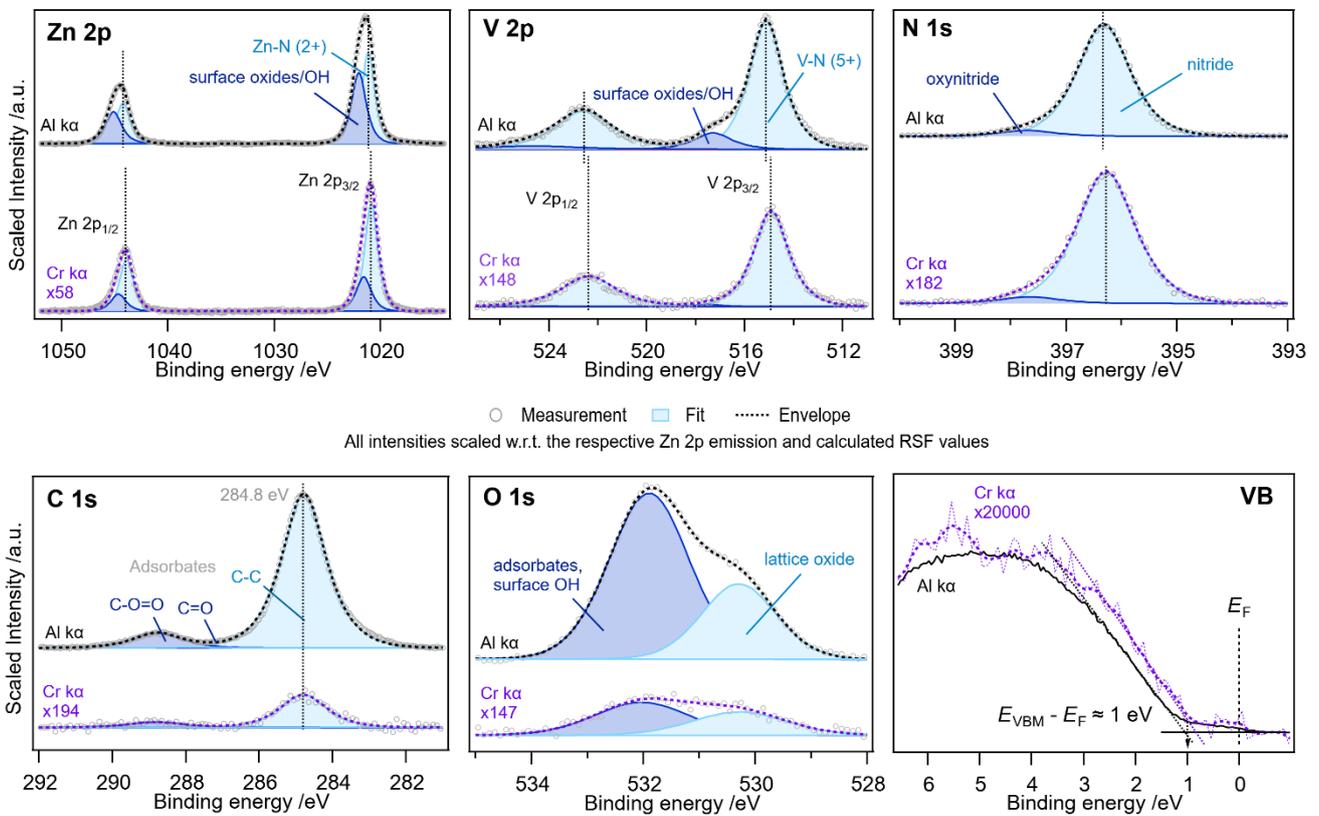

**Figure 7**. XPS and HAXPES characterization of ex-situ treated $Zn_2VN_3$ thin films on glass. Shown are the core level spectra of Zn 2p, V 2p, N 1s, C 1s, O 1s as well as the valence band region for measurements performed with Al kα (XPS) as well as Cr kα (HAXPES) radiation. The information depth ranges from 3 nm (Zn 2p) to 6 nm (VB) for the XPS, as well as 20 nm (Zn 2p) to 24 nm (VB) for the HAXPES measurement. The higher information depth of HAXPES allows analyzing the nitride despite the presence of surface oxide- and adsorbate layers. Cr kα count rates are scaled to account for the lower photoionization cross-section of HAXPES at lower binding energies.

**Figure 7** shows a complete set of core level spectra as well as the valence band emission for a sputtered $Zn_2VN_3$ thin film on glass. In the XPS measurement, both Zn 2p as well as V 2p core level spectra clearly show the presence of a shoulder at higher binding energies which can be attributed to surface oxidation. For the HAXPES measurements, the intensity of this component is reduced significantly, indicating that the oxidation does not extend into the bulk of the material. This was further confirmed by performing a sputter depth profile (see **Figure S5**), which showed oxygen concentrations of < 3 at.% in the bulk. The Zn 2p core level emission typically shows only minimal shifts for different oxidation states. The associated Zn LMM Auger electron emission, on the other hand, is much more sensitive to changes in the local chemical environment and allows us to clearly differentiate between the surface oxide and the underlying nitride.[44] **Figure S6** shows a Wagner plot for the Zn 2p as well as the Zn LMM Auger electron emission. For the V 2p core level shifts in binding energy are much more pronounced. Here, vanadium oxides of different oxidation states are clearly separated in binding energy.[44] Consequently, the comparatively narrow V $2p_{3/2}$ peak observed in the HAXPES measurements of $Zn_2VN_3$ indicates that V predominantly exhibits a single oxidation state, i.e. $V^{5+}$. As more V is added to $Zn_2VN_3$ and the material becomes V-rich, the V $2p_{3/2}$ peak develops a shoulder at lower binding energies indicating the presence of $V^{4+}$ as well as $V^{5+}$ due to isostructural alloying on the cation site. For $Z_{1-x}V_xN$ films with x=0.52 we observe mostly a $V^{5+}$ oxidation state (see **Figure S7**). This is in good agreement with our results from XRD mapping, which show an extended region of stability of a wurtzite $Z_{1-x}V_xN$ phase with an onset of decomposition close to x=0.5 and x=0.4 for lower and higher deposition temperatures, respectively (see **Figure 3**). Using the higher information depth of the HAXPES measurement, it is also possible to probe the position of the valence band maximum (VBM). For phase pure $Zn_2VN_3$, the VBM is located at approximately 1 eV ± 0.1 eV below the Fermi level, indicating only a weak p-type doping in $Zn_2VN_3$ in contrast to the predictions from DFT calculations. The difference in position of Fermi levels for simulated and experimentally derived $Zn_2VN_3$ may be attributed to the structural defects in the synthesized thin film. **Table 1** shows the core level peak binding energies ($CL_{BE}$), Auger electron kinetic energies ($AE_{KE}$), as well as modified Auger parameter (AP) for $Zn_2VN_3$ (the Auger parameter is defined as the sum of $CL_{BE}$ and $AE_{KE}$). The AP is insensitive to charging effects and facilitates the comparison of XPS results from different instruments.[43,45,46] By providing this data we aim to supply a reference value, which will be useful in identifying $Zn_2VN_3$ among other phases by other groups in the future.

**Table 1**: Core level binding energies $CL_{BE}$ as well as Auger electron kinetic energies $AE_{KE}$ for $Zn_2VN_3$. The modified Auger parameter is calculated as AP= $CL_{BE}$ + $AE_{KE}$.

| Core level | $CL_{BE}$ /eV | ($2p_{3/2}$–$2p_{1/2}$) /eV | Auger line | $AE_{KE}$ /eV | AP /eV |
|---|---|---|---|---|---|
| Zn $2p_{3/2}$ | 1021.2 ± 0.1 | 23.08 | Zn $L_3M_{45}M_{45}$ | 990.5 ± 0.1 | **2011.7 ± 0.2** |
| V $2p_{3/2}$ | 515.1 ± 0.1 | 7.45 | V $L_3M_{45}M_{45}$ | 510.4 ± 0.2 | **1052.5 ± 0.3** |
| N 1s | 396.3 ± 0.1 | - | N KLL | 384.5 ± 0.5 | **780.8 ± 0.6** |

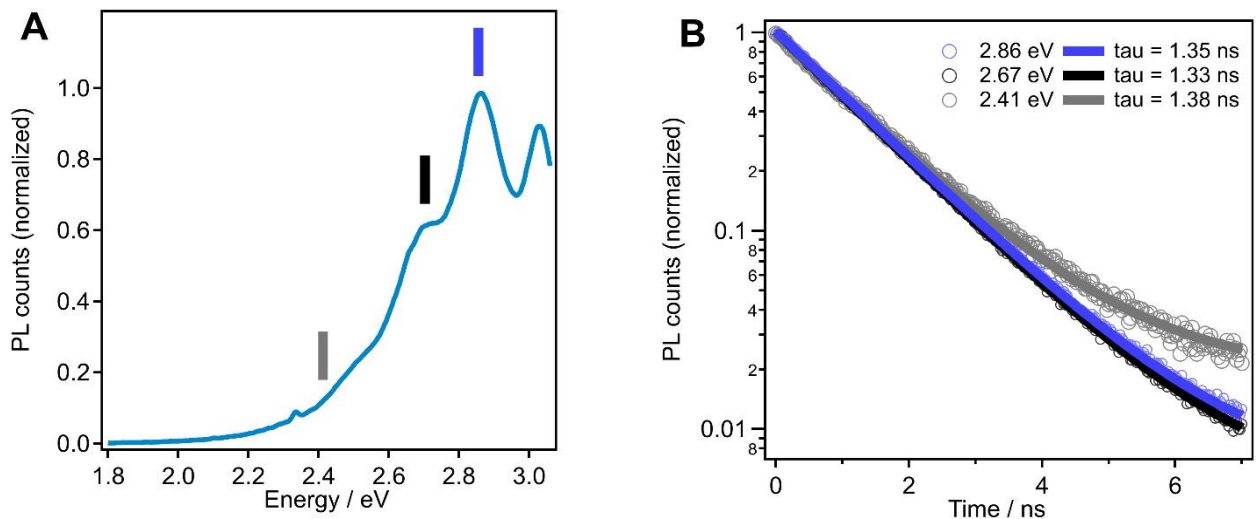

**Figure 8**. Room-temperature PL spectrum and PL lifetime of a $Zn_2VN_3$ thin film. (a) PL spectrum upon photoexcitation at 355 nm. (b) PL kinetics at 2.86 eV, 2.67 eV, and 2.41 eV, indicated in (a) *via* the three vertical markers in blue, black, and grey, respectively; all PL bands exhibit close to mono-exponential decay with a lifetime of 1.3-1.4 ns.

To explore the potential of our newly synthesized $Zn_2VN_3$ thin films for optoelectronic applications, we studied their photoluminescence (PL) characteristics following photoexcitation at 355 nm. **Figure 8a** displays the room-temperature PL spectrum of a $Zn_2VN_3$ thin film on a glass substrate, featuring strong and broadband emission throughout the visible range, with three clearly discernible PL peaks at ~2.5 eV, ~2.67 eV, and ~2.86 eV, respectively. Furthermore, a broad tail extends up to slightly beyond 2 eV. Overall, these emission energies are in line with the band structure calculations in **Figure 2c**. Importantly, the observation of PL already at room temperature is encouraging for a variety of applications, *e.g.* in optoelectronic or photocatalytic devices, as it points to a favorable electronic landscape with either a limited amount of defects or else electronically benign defects. Second, the still poorly explored Zn-V-N phase space opens the possibility for further tuning the PL characteristics *via* (compositional) materials engineering. For example, while we have not yet attempted to investigate the origin of the three observed PL bands, it is worth noting a recent related work on $ZnGeN_2$ [47] where the authors ascribed a comparable progression of PL bands in the visible, spaced by about 0.2-0.3 eV, to $Zn_{Ge}$-$Ge_{Zn}$ antisite defect complexes. Given the similarity in these spectral characteristics of $ZnGeN_2$ and $Zn_2VN_3$, as well as evidence of cation-disorder from XRD analysis it is reasonable to conclude that a similar effect is observed in $Zn_2VN_3$. To investigate the effect of the structure on the electronic properties of the material, band-structure calculations were performed on $Zn_2VN_3$ using both an ordered orthorhombic as well as a cation-disordered wurtzite structure (see **Figure S8**). The results clearly indicate a closing of the band gap with the introduction of cation-disorder to the material, which is in good agreement with the results from optical analysis. While the absolute band gap values of the GGA method used in this calculation are less accurate compared to the HSE calculations presented in **Figure 2**, the general trend is expected to be maintained. We suggest that future studies may explore and potentially exploit the effect of such structural changes as means to further engineer the desired PL characteristics.

**Figure 8b** displays the PL kinetics following pulsed UV excitation at 355 nm. A close-to-mono-exponential decay with a lifetime of 1.35 ns is observed for the main PL peak at 2.85 eV, and *quasi*-identical lifetimes also for the lower-energy bands at 2.67 eV and 2.41 eV, respectively. While more studies are needed to elucidate the decay mechanism, the nearly mono-exponential decay points to a relatively simple decay scheme with a single dominant rate constant. Overall, these findings are also supported by photospectrometry measurements. **Figure S9** shows transmittance and reflectance spectra of ~900 nm thick $Zn_2VN_3$ thin films deposited on glass substrate. The transmittance edge is found to be at ~2.15 eV, consistent with the revealed tail towards 2 eV in the PL spectra. In addition, significant sub-bandgap absorption in $Zn_2VN_3$ thin film is observed, possibly related to said cation-disorder. Further, more detailed studies on defects in $Zn_2VN_3$ are carried out and will be reported in the future.

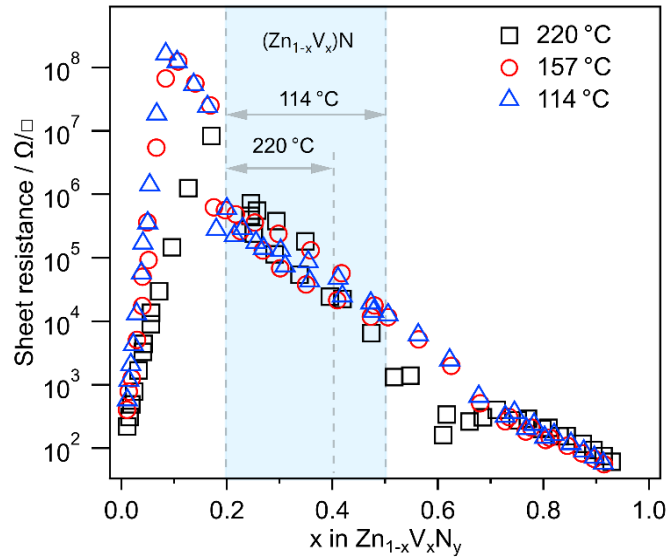

**Figure 9**. Sheet resistance of 300 nm – 500 nm thick Zn-V-N thin films synthesized at substrate temperatures of 114 °C (blue triangles), 157 °C (red circles) and 220 °C (black squares). The blue region represents the approximate phase boundaries from the XRD screening.

Electrical properties of Zn-V-N thin films were measured using the four-point probe method. The resistivity of stoichiometric $Zn_2VN_3$ thin film was found to be ~1.16 Ωcm, which corresponds to the semiconducting range. These results are confirmed by temperature dependent transport measurements on $Zn_2VN_3$ films on sapphire. The results of this analysis are shown in **Figure S10**. The films show p-type conductivity with a carrier concentration on the order of $1\times10^{17}/cm^3$. While the carrier concentration remains relatively stable with temperature, an increase of mobility is observed with temperature, which could be indicative of a hopping-transport behavior. To investigate the effect of the composition on the electrical properties a sheet resistance mapping was performed. **Figure 9** shows sheet resistance ($R_s$) mapping of 300 nm – 500 nm thick Zn-V-N thin films synthesized at substrate temperatures of 114 °C, 157 °C, and 220 °C, respectively. The composition region attributed to orthorhombic $Zn_2VN_3$ and $ZnVN_2$ phases is characterized by $R_s$ in a wide range from $10^4$ Ω/□ to $10^6$ Ω/□. Note that $R_s$ of the $Zn_{1-x}V_xN$ films decreases steadily with increasing V atomic content from about *x*

= 0.2 to $x$ = 0.5 for samples deposited at 114 °C and 157 °C. This evidences that electrical properties of $Zn_2VN_3$ thin films can be controlled by tuning the Zn/V cation ratio. On the other hand, a clear drop in $R_s$ is observed for the samples synthesized at 220 °C for V concentrations higher than $x$ = 0.4. This can be explained by the decomposition of the $Zn_{1-x}V_xN$ phase at the high synthesis temperature and the associated formation of a highly conductive precipitates such as VN (see **Figure 3a**). In addition, these films exhibit metallic-like reflectance spectra and low transparency (data not shown in the manuscript) further supporting the presence of a VN secondary phase owing to the narrow thermodynamic stability region of this metastable phase. Further studies are being conducted to increase the non-equilibrium solubility to higher V contents and to fabricate phase-pure $ZnVN_2$ thin films in the future.

## 4. Conclusion

Using a combination of computational prediction and combinatorial PVD the Zn-V-N phase space was explored to perform a rapid development of new functional nitrides for optoelectronic applications. This screening resulted in the synthesis of the novel $Zn_2VN_3$ phase by reactive co-sputtering in a mixed plasma of Ar and $N_2$. Moreover, the novel phase was isolated and synthesized as a single-phase film on a substrate of 50.8 mm x 50.8 mm size. The deposition of $Zn_2VN_3$ on (001) sapphire substrates resulted in improved crystallinity and epitaxial growth of the film with (002) out-of-plane orientation. Strikingly the films exhibit 6-fold rotational symmetry. This can be attributed to cation-disorder resulting in a wurtzite structure of the material. RBS/ERDA characterization demonstrates a stoichiometric composition of $Zn_2VN_3$ with a negligible amount of C and O in the bulk of the film. Furthermore, the films are stable in ambient conditions as demonstrated by HAXPES and STEM analysis on ex-situ specimens. We find that $Zn_2VN_3$ is a weakly p-doped wide bandgap semiconductor with carrier concentration on the order of $1\times10^{17}/cm^3$. The material exhibits broadband PL at room temperature across the visible range, in good agreement with the direct bandgap of 2.35 eV predicted by DFT calculations. In addition, the resistivity can be tuned over a wide range by varying the V/Zn cation ratio, making this an interesting material for optoelectronic applications. On the other hand, a significant amount of sub-bandgap absorption and relatively short PL lifetime of about 1.3 ns both point towards the presence of electronically active defects that should be addressed to further improve the optoelectronic properties of this novel compound.

**Supplementary Materials**

The supplementary materials to this work contain the following information: XRD patterns of 45 $Zn_2VN_3$ thin film samples deposited on the glass substrate of 50.8 mm x 50.8 mm size; composition of $Zn_2VN_3$ thin film measured using RBS and ERDA; a detailed description of the XPS/HAXPES measurements; XPS-Sputter depth profile; HAXPES measurement of V 2p core level spectrum for $Zn_{1-x}V_xN$ thin films with V content of $x$ = 0.33 and $x$ = 0.52; Wagner plot for the modified Auger parameter α' of Zn 2p and Zn LMM for selected compounds; transmittance and reflectance spectra of $Zn_2VN_3$ thin films as well as electronic transport measurements. In addition, we provide the .cif files for the refined structures from GGA calculations for orthorhombic $Zn_2VN_3$ and $ZnVN_2$ as well as the experimentally refined wurtzite $Zn_2VN_3$.

**Author contributions**

S.Z. and S.S wrote the manuscript with contributions from S.C.B. and M.V.K. S.S. conceived the project and supervised the research work. S.Z. synthesized samples and performed XRD, XPS, UV-Vis-NIR and electrical characterization. A.A.K. carried out density functional theory calculations. S.C.B. measured PL and TRPL spectra as well analyzed the results. M.S. prepared FIB lamella, N.O. performed STEM characterization and analyzed the results together with S.S. S.S. performed XPS/HAXPES measurements and analyzed results. F.L. performed the electronic transport measurements. All co-authors reviewed and commented on the manuscript.

**Acknowledgements**

S.Z. acknowledges funding from the EMPA internal research call 2020. Financial support from the Swiss National Science Foundation (R'Equip program, Proposal No. 206021_182987) is gratefully acknowledged. K.A.A. acknowledges funding from the European Research Council (ERC) under the European Union's Horizon 2020 research and innovation program (grant agreement No. 101002219).  M.V.K. and S.C.B. acknowledge funding by the European Union's Horizon 2020 program, through a FET Open research and innovation action under the grant agreement no. 899141 (PoLLoC). The authors would like to acknowledge Jan Sommerhäuser for his support with the four point probe measurement setup, Christof Vockenhuber and Max Döbeli for their help with RBS and ERDA characterization, Thomas Amelal for his assistance with the sputter deposition as well as Raman Palikarpau and Aliaksandr Khinevich for their support with machine learning predictions (not included in the manuscript). The authors also acknowledge CSC – IT Center for Science, Finland, for computational resources.

# Supplementary Information:

# Synthesis and Characterization of the Ternary Nitride Semiconductor $Zn_2VN_3$: Theoretical Prediction, Combinatorial Screening and Epitaxial Stabilization

Siarhei Zhuk[a], Andrey A. Kistanov[b], Simon C. Boehme[a,c], Noémie Ott[a], Fabio La Mattina[a], Michael Stiefel[a], Maksym V. Kovalenko[a,c], Sebastian Siol[a,*]

a) Empa – Swiss Federal Laboratories for Materials Science and Technology,
8600 Dübendorf, Switzerland

b) Nano and Molecular Systems Research Unit, University of Oulu,
90014 Oulu, Finland

c) Laboratory of Inorganic Chemistry, Department of Chemistry and Applied Biosciences, ETH Zürich,
8093 Zürich, Switzerland

*E-mail: sebastian.siol@empa.ch

# 1. Methods

## 1.1    Synthesis of combinatorial sample library

1.1 mm thick borosilicate glass substrates of 50.8 mm x 50.8 mm size were ultrasonically cleaned in acetone and ethanol for 5 min each before loading into the deposition chamber of AJA 1500F sputtering system. Prior to the deposition, the chamber was evacuated to the pressure of below $3\cdot10^{-6}$ Pa using a turbomolecular pump. The thin films were synthesized by reactive radio-frequency (RF) co-sputtering of Zn (99.995% purity, 2" diameter) and V (99.9% purity, 2" diameter) metal targets in mixed plasma of Ar and $N_2$. The deposition pressure was maintained at 1 Pa while Ar and $N_2$ flows were 12 sccm and 18 sccm, respectively. $N_2$ gas was introduced directly into the chimneys of the sputter guns, facilitating dissociation and ionization of the reactive gas and thus increasing its chemical potential. The sputtering powers in the range 16-28 W for Zn and 30-150 W for V target were used to synthesize Zn-V-N thin films of various cation ratios. It is worth mentioning that sputtering powers on both guns were adjusted to maintain the average thin film deposition rate constant.

In order to fabricate ternary metal nitride thin film libraries for the combinatorial analysis, the composition and temperature gradients were applied during the deposition process. Sputtering was carried out at incident angles of approximately 30° without substrate rotation, resulting in a compositional gradient across the library. Furthermore, the substrate was only partially clamped to the heated sample holder resulting in an orthogonal temperature gradient in the range from 220 °C to 114 °C. After the deposition, the sample was cooled down in process gas.

## 2. Results and discussion.

### 2.1 Structural Analysis

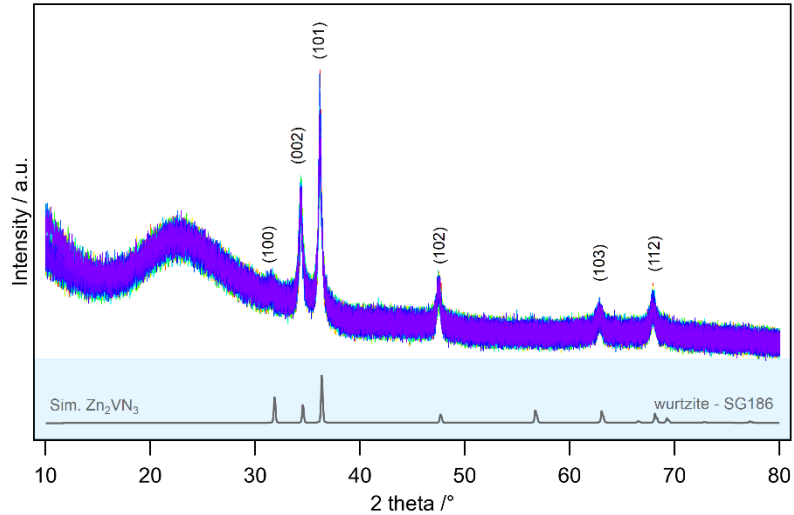

**Figure S1**. XRD patterns of 45 Zn$_2$VN$_3$ thin film samples deposited on glass substrate of 50.8 mm x 50.8 mm.

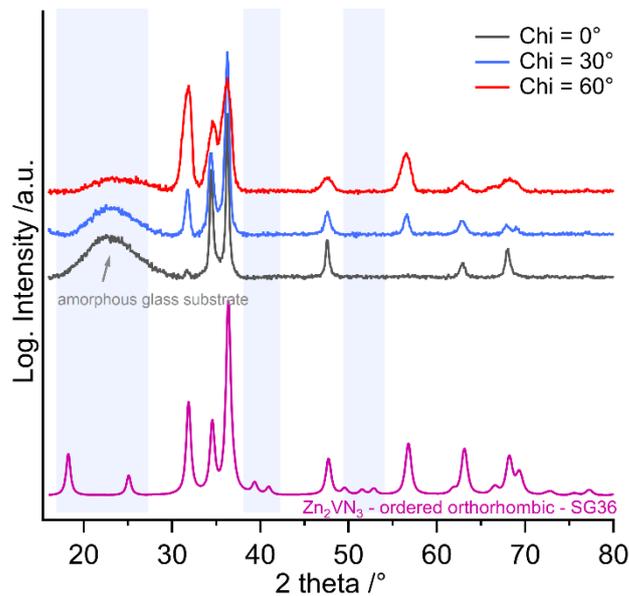

**Figure S2**. XRD patterns of Zn$_2$VN$_3$ thin films measured at different tilts of the glass substrate. The highlighted regions show expected peaks of the orthorhombic XRD pattern, which are not observed in the measurement regardless of the tilt angle.

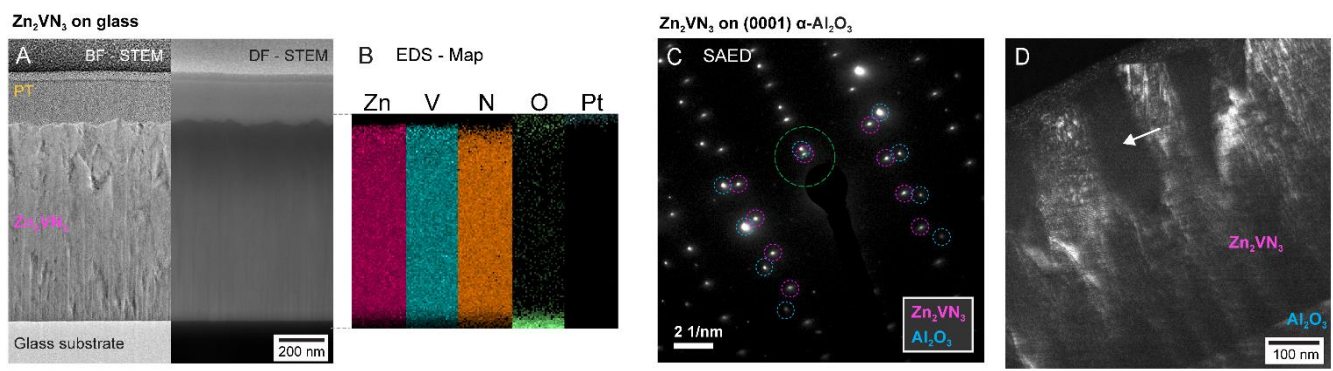

**Figure S3**. Microstructural analysis of $Zn_2VN_3$: Shown are (a) bright field (BF) and dark (DF) field STEM micrographs as well as (b) representative EDS maps of the film forming elements for layers on glass. No composition fluctuations are observed in the EDS maps, indicating a homogenous distribution of the film-forming elements. (c) Selected area electron diffraction (SAED) for epitaxial $Zn_2VN_3$ on $Al_2O_3$, (d) SA DF-STEM micrograph acquired using the peaks highlighted in green in the SAED pattern. Dark spots indicate grains of different orientation.

## 2.2 RBS and ERDA analysis

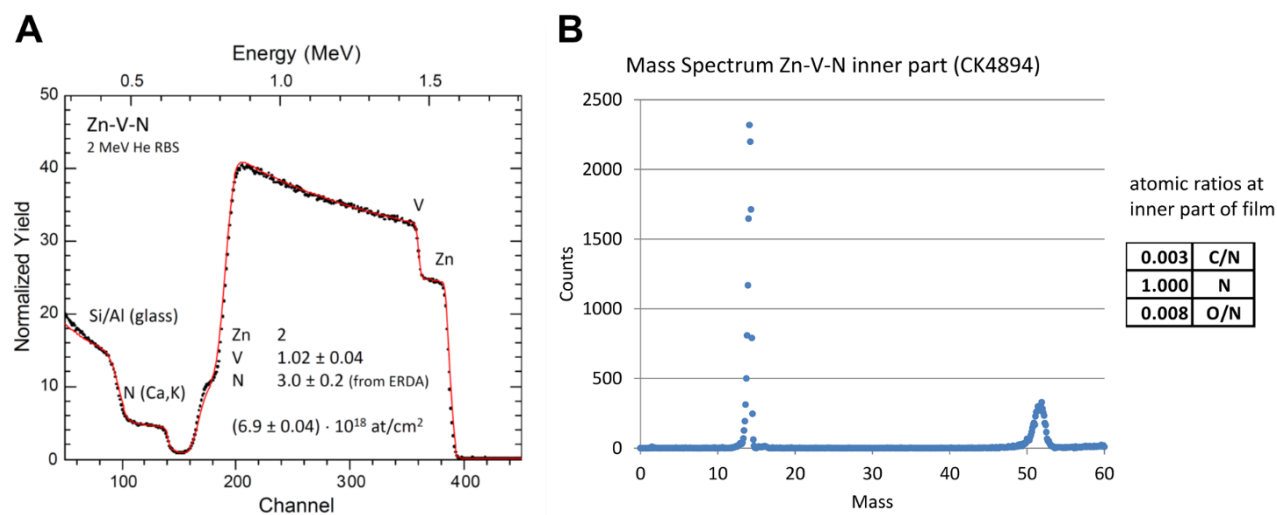

**Figure S4**. Composition of $Zn_2VN_3$ thin film measured using (a) RBS and (b) ERDA.

## 2.3 XPS/HAXPES Analysis

### 2.3.1 Probing depth of the measurements

The excitation with X-rays of different photon energies leads to a variation in information depth in XPS and HAXPES measurements. In addition, it is important to consider, that the kinetic energy of the photoelectrons varies significantly depending on the binding energy of the probed core level. The kinetic energy of Auger electrons follows an interatomic relaxation process and hence is independent of the excitation source. The inelastic mean free path λ (IMFP) of photoelectrons in a solid can be estimated using the Tanuma, Powell, Penn formula (TPP2M).[46,47] In this work we used the Quases software package for the calculation of the IMFP in $Zn_2VN_3$. For this calculation we assumed a density of 4.74 g/cm$^3$ as determined by RBS measurements. At a take-off angle α the probing depth can be calculated via 3 × λ cos(α). Table S1 shows the binding energies (BE), kinetic energies (KE) and probing depth for the regions relevant to this work.

**Table S1:** Relevant regions measured in XPS (Al kα) /HAXPES (Cr kα) analyses as well as their respective binding energies (BE), kinetic energies (KE), inelastic mean free path (IMPF) and their probing depth at an electron take of angle of 5° to the substrate normal.

| Region | BE /eV | KE /eV | | IMFP/Å | | 3*IMFP*COS(5°) /nm | |
|---|---|---|---|---|---|---|---|
| | | Al kα | Cr kα | Al kα | Cr kα | Al kα | Cr kα |
| **Vb** | 0 | 1486.7 | 5417.7 | 28.4 | 81.4 | 8.5 | 24.3 |
| **Zn 2p** | 1021 | 465.7 | 4396.7 | 12.0 | 68.1 | 3.6 | 20.3 |
| **V 2p/O 1s** | 528 | 958.7 | 4889.7 | 20.3 | 74.4 | 6.1 | 22.2 |
| **N 1s** | 396 | 1090.7 | 5021.7 | 22.3 | 76.0 | 6.7 | 22.7 |
| **V LMM** | | 510.0 | 510.0 | 12.7 | 12.7 | 3.8 | 3.8 |
| **Zn LMM** | | 990.5 | 990.5 | 20.8 | 20.8 | 6.2 | 6.2 |

### 2.3.2 Sputter depth profile and oxygen contamination

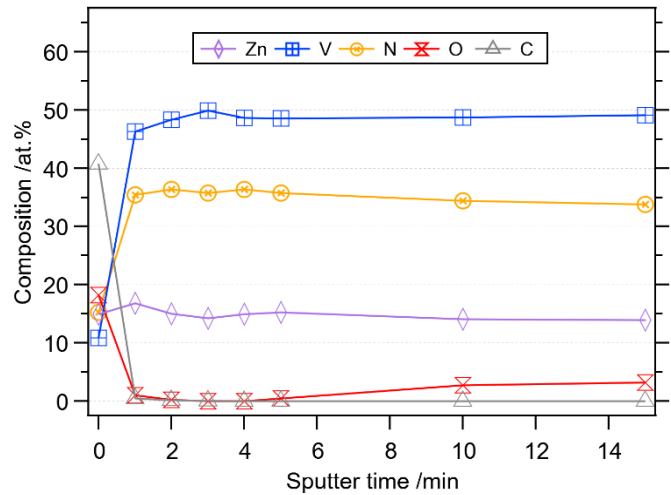

**Figure S5**. XPS-Sputter depth profile: O and C signals quickly decrease as the surface layer is removed. In addition, preferential sputtering leads to an immediate decrease in Zn/V ratio after the first sputter cycle. The semi-quantitative composition analysis indicates a bulk oxygen level of less than 3 at.%.

### 2.3.3 Chemical state analysis

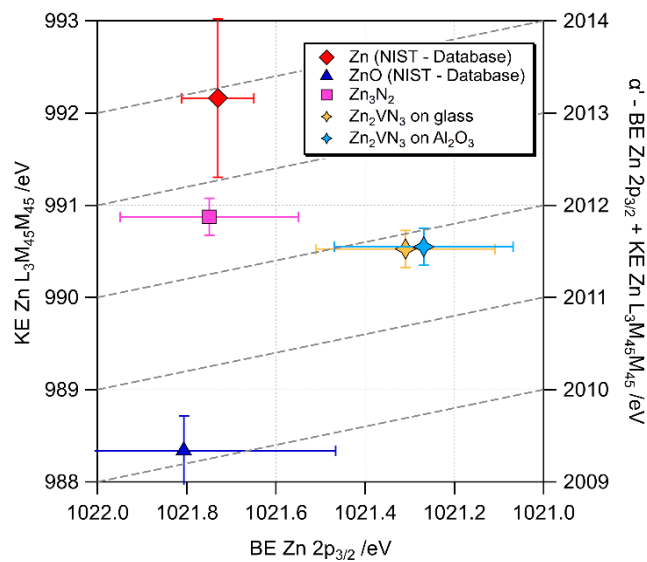

**Figure S6**. Wagner plot for the modified Auger parameter α' of Zn 2p and Zn LMM for selected compounds. Zn and ZnO data are taken from the NIST XPS database. $Zn_3N_2$ and $Zn_2VN_3$ were measured as part of this work. In contrast to analysis of the Zn 2p core level binding energy alone, the Auger parameter allows for a clear distinction of the respective phases.

## 2.3.4 Determination of the V oxidation state

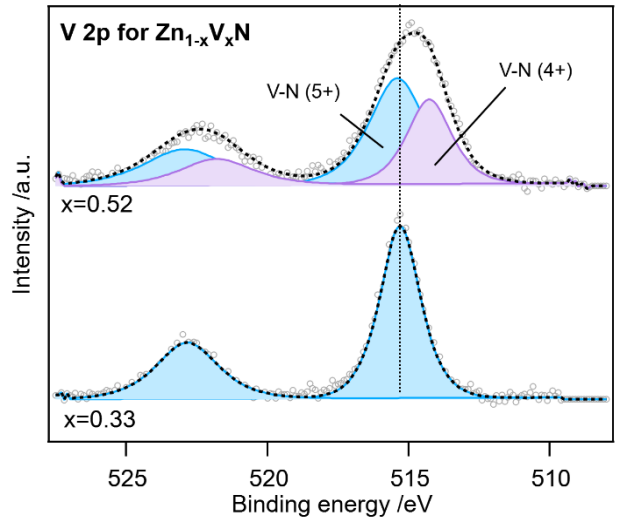

**Figure S7**. HAXPES measurement of the V 2p core level spectrum for $Zn_{1-x}V_xN$ thin films grown on $Al_2O_3$ substrates at a deposition temperature of approximately 114 °C. For stoichiometric $Zn_2VN_3$ (x = 0.33) only one oxidation state $V^{5+}$ is observed. For higher vanadium concentrations (x = 0.52) a second component is observed indicating a mixed oxidation state of $V^{5+}$ and $V^{4+}$.

## 2.4 DFT calculations

In order to further investigate the effect of cation disorder on the properties of $Zn_2VN_3$, additional DFT calculations were carried out. The wurtzite-like structure of $Zn_2VN_3$ was created based on the ZnN wurtzite structure, where Zn atoms were randomly substituted with V atoms so that the V/Zn ratio was equal to 1/2. Energy band diagrams for ordered orthorhombic and disordered wurtzite-like $Zn_2VN_3$ structures are shown in **Figures S8**. Due to a high computational demand of the HSE method, the electronic structure of the ordered orthorhombic and disordered wurtzite-like $Zn_2VN_3$ were calculated using the GGA method. It has been revealed that the transition from an ordered orthorhombic $Zn_2VN_3$ to a wurtzite structure with random distribution of cations is characterized by ~5-folds decrease of the band gap value. This phenomenon is consistent with the experimentally observed sub-bandgap absorption (see **Figure S9**) and tail in PL spectra (see **Figure 8a**) demonstrating detrimental effect of cation disorder on the optical properties of $Zn_2VN_3$ thin films.

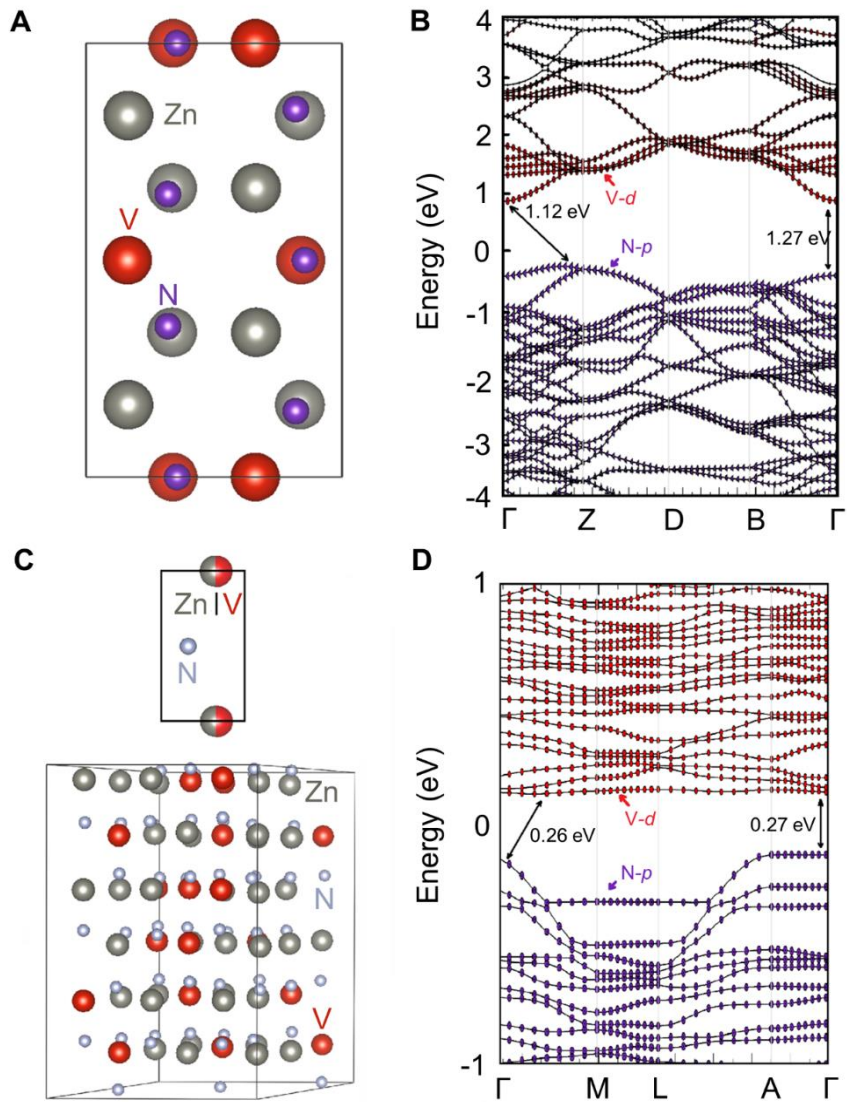

**Fig. S8.** (a) The unit cell of ordered orthorhombic $Zn_2VN_3$ and (c) the unit cell and calculated supercell of wurtzite-like $Zn_2VN_3$ with random distribution of cations. Energy band diagrams calculated using GGA method for (b) ordered orthorhombic $Zn_2VN_3$ and (d) wurtzite-like $Zn_2VN_3$.

## 2.5 Optical measurements

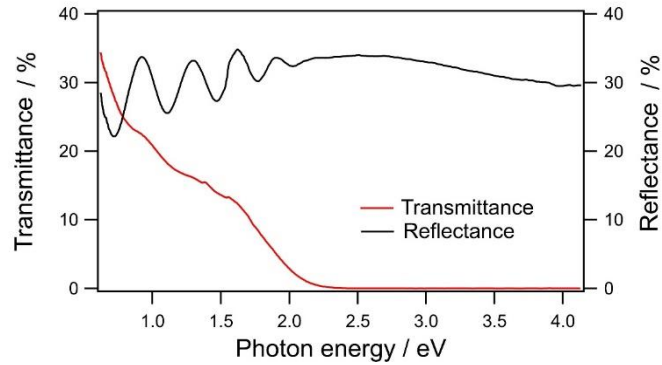

**Figure S9**. Transmittance and reflectance spectra of Zn$_2$VN$_3$ thin films on glass.

## 2.6 Hall effect measurements

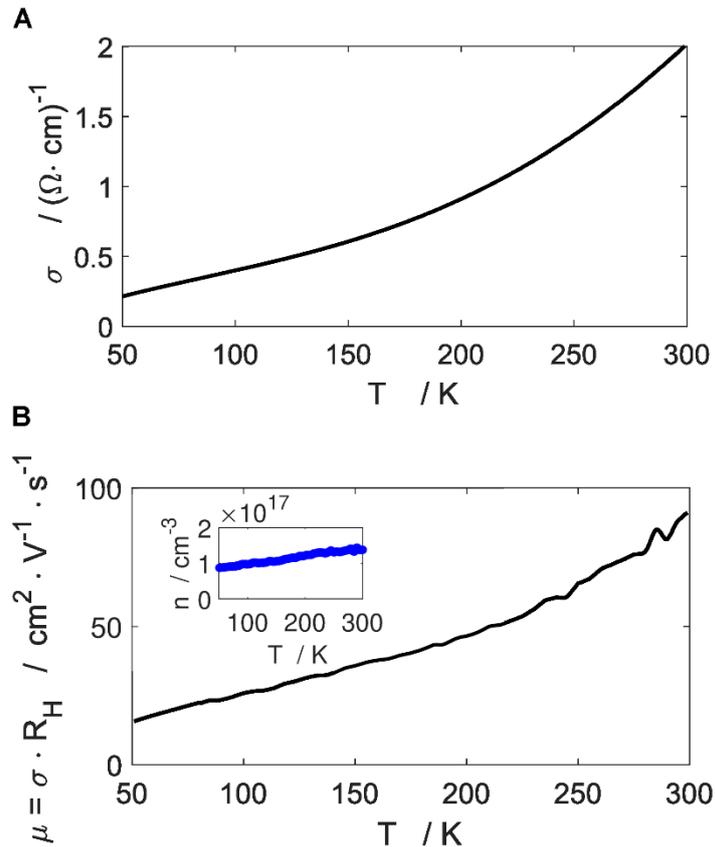

**Figure S10.** (a) Conductivity and (b) Mobility versus temperature in a single phase film of Zn$_2$VN$_3$ (560 nm thick) on a sapphire substrate. The total mobility was obtained as product of Hall coefficient times the conductivity. Carrier concentration decreases of about 30% at low temperature as shown in the inset of panel (b). The Hall coefficient was positive over the full temperature scan, indicating p-type of the majority of carriers.

A Zn$_2$VN$_3$ film of a thickness $d$=560 nm deposited on sapphire was used for transport measurements. A mechanical scribing confines these measurements to a region of about 10 x 5.01 mm$^2$. The conductivity and Hall coefficient were acquired in typical Hall-bar geometry with four-probe schema. The conductivity and Hall coefficient were acquired in typical Hall-bar geometry with four-probe schema. The Hall effect signal $V_H$ appears at the edges of the Hall bar when a magnetic field is applied. However the measured voltage $(V_+-V_-)$ is the sum of the transverse Hall voltage ($V_H$) plus the longitudinal voltage drop due to the sample resistance $V_r$ (along $\Delta x$ in Figure S11a). These terms can be separated because only $V_H$ depends by the magnetic field. The Hall coefficient is calculated as $R_H = V_H \cdot d \cdot (I \cdot B)^{-1}$, with $d$ the sample thickness, $I$ the current and $B$ the magnetic field. We performed measurements at constant $I$ and multiple $B$ to determine $R_H$ in this measurement geometry. The conductivity (Figure S11b) was obtained using a four point probe method by measuring the voltage drop $V_+-V_-$ under a constant current of 0.1 µA during the temperature scan. The conductivity was calculated as $\sigma = I \cdot \Delta x \cdot ((V_+-V_-) \cdot L \cdot d)^{-1}$ from the sample geometry. Here $d$ is the film thickness, $L$ is the lateral size of the film and $\Delta x$ is the longitudinal distance between the electrical contacts across which we measure the voltage drop.

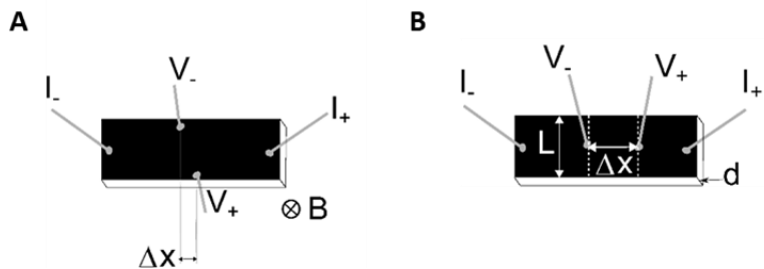

**Figure S11.** Contact geometry for the electrical transport measurements: a) Hall effect measurements, b) Conductivity measurements.